\documentclass[]{interact}

\usepackage{color}
\usepackage{epstopdf}
\usepackage[caption=false]{subfig}

\usepackage[numbers,sort&compress]{natbib}% Citation support using natbib.sty
\bibpunct[, ]{[}{]}{,}{n}{,}{,}% Citation support using natbib.sty
\makeatletter% @ becomes a letter
\def\NAT@def@citea{\def\@citea{\NAT@separator}}% Suppress spaces between citations using natbib.sty
\makeatother% @ becomes a symbol again

\theoremstyle{plain}% Theorem-like structures provided by amsthm.sty

\theoremstyle{definition}

\theoremstyle{remark}

\begin{document}

% \articletype{ARTICLE TEMPLATE}% Specify the article type or omit as appropriate

\title{
Improved cutoff functions for
short-range potentials and the Wolf summation
% Coulomb interactions in neutral, unpolarized systems
}

\author{
\name{Martin H. M\"user, Dept. of Materials Science and Engineering, Saarland University, Saarbr\"ucken, Germany}
}

\maketitle

\begin{abstract}
A class of radial, polynomial cutoff functions $f_{\textrm{c}n}(r)$ for short-ranged pair potentials or related expressions is proposed.
Their derivatives  up to order $n$ and $n+1$ vanish at the outer cutoff $r_\textrm{c}$ and an inner radius $r_\textrm{i}$, respectively.
Moreover, $f_{\textrm{c}n}(r \le r_\textrm{i}) = 1$ and 
$f_{\textrm{c}n}(r\ge r_\textrm{c})=0$.
It is shown that the used order $n$ can qualitatively affect results:
stress and bulk moduli of ideal crystals are unavoidably discontinuous with density for $n=0$ and $n=1$, respectively. 
Systematic errors on energies and computing times decrease by approximately 25\% for Lennard-Jones  with $n=2$ or $n=3$ compared to standard cutting procedures. 
% $r_\textrm{c}$ to be reduced while improving accuracy compared to standard cutoff schemes. 
% Thus, $n=2$ and $n=3$ appear to be good general compromises between reproducing cohesive energies closely while avoiding cutoff artifacts due to large forces and stiffnesses. % at $r \lesssim r_\textrm{c}$. 
%
Another cutoff function turns out beneficial to compute Coulomb interactions using the Wolf summation, which is shown to not properly converge when local charge neutrality is obeyed only in a stochastic sense.  
However, for all investigated homogeneous systems with thermal noise (ionic crystals and liquids), the modified Wolf summation, despite being infinitely differentiable at $r_\textrm{c}$, converges similarly quickly as the original summation.
Finally, it is discussed how to reduce the computational cost of numerically exact Monte Carlo simulations using the Wolf summation even when it does not properly converge.  
\end{abstract}


\section{Introduction}

The efficiency of molecular simulations hinges on the truncation of interaction potentials~\cite{Allen2017,Frenkel2002}.
One possibility to achieve that is to multiply the interaction potential, or, functions entering their calculation, with cutoff functions~\cite{Stillinger1985PRB,Tersoff1986PRL,Steinbach1994JCC}. 
However, there are two major, mutually exclusive requirements on them.
The cutoff radius $r_\textrm{c}$ should be as large as possible to reduce systematic errors~\cite{Patra2003BJ} but also be as small as possible to boost computational efficiency.
Similarly, given a particular value for $r_\textrm{c}$,  $f_\textrm{c}(r)$ should be close to unity for as long as possible to reduce discrepancies from the real energy. 
At the same time, $f_\textrm{c}(r)$ should be decreased to zero as smoothly as possible to avoid freakishly large forces or curvatures at large distances, which generally induce undesired behavior~\cite{Mattoni2005PRL,Pastewka2008PRB}.  
A compromise is certainly needed but it does not seem that a generally applicable one has been identified. 

Another common strategy to cut off short-range potentials is by making all their derivatives up to $n$'th order vanish continuously at $r_\textrm{c}$ through
\begin{equation}
U_{\textrm{s}n}(r) = \left\{U(r)-\mathcal{T}[U(r),r_\textrm{c},n] \right\}\Theta(r_\textrm{c}-r),
\end{equation}
where $U(r)$ is a pair-potential or a related local function,
$\mathcal{T}[U(r),a,n]$ the $n$'th order Taylor expansion of $U(r)$ about $r = a$, and $\Theta(r)$ the Heaviside function.
A shifted-potential (SP) potential is obtained for $n=0$, a shifted-force (SF) potential for $n=1$~\cite{Beck2005B,Waibel2019JCTC}, and a shifted-curvature (SC) potential for $n=2$.
A disadvantage of the shifting procedure is that binding or cohesive energies decrease rather quickly with $n$ at fixed $r_\textrm{c}$. 
For this but also for other reasons, it is often desirable to sum up potentials or other local functions so that the contribution from nearest neighbors is exact, for example, when computing the embedding density within a potential based on the embedded-atom method~\cite{Daw1984PRB}.
To achieve this, the partial densities are mutliplied with cutoff functions, $f_\textrm{c}(r)$, which are set to unity up to a an inner radius ${r}_\textrm{i}$ and then swiftly decreased toward zero~\cite{Tersoff1986PRL}.
However, in order to avoid qualitative cutoff artifacts, $f_\textrm{c}(r)$ has to approach zero in a sufficiently smooth fashion.

Good cutoff functions are central to balance computational efficiency and systematic errors but are surprisingly little discussed even in stellar text books on molecular simulation~\cite{Allen2017,Frenkel2002}. 
Unfortunately, there is no unique optimum.
It would depend not only on the potential but also on the property of interest. 
For example, when studying sublimation, reproducing the energies themselves is crucial.
However, forces and curvature of potentials determine mechanical properties.
In this context, it is useful to keep in mind that an attractive potential that was cut without shifting leads to a diverging force at $r_\textrm{c}$ so that a corresponding bond cannot be broken with a finite force. 
In a cut-and-shift potential, stress is still discontinuous in density for ideal crystals (and thus potentially for other systems too) so that upon reversion the density is discontinuous in pressure. %
Each higher order in a cut-and-shift procedure mitigates artifact to one higher-order in the response function so that $n=3$ is the lowest cut-and-shift order, which systematically avoids a discontinuity of elastic properties with pressure. 
% Even a shifted-force potential ($n=2$) can still induce artifacts, e.g., a discontinuous changes of elastic properties with density. 
%
% If you're an Academic, and chances are high for that or you would not be reading this paper, you may want to add the argument that symplectic integrators require not only forces but also curvatures to be continuous.
%
% It thus seems as though the lowest generally acceptable cutoff function should satisfy $n \ge 3$. 
%
% Even higher orders risk to do poorly on binding energies or to necessitate large cutoff but small inner radii. 

In this paper, cutoff functions with beneficial properties are proposed. 
% are proposed which should generally outperform  commonly used.
%
The cutoff functions are designed to take a value of unity up to $r_\textrm{i}$ and to approach zero continuously as their argument approaches $r_\textrm{c}$. 
% , which should not be smaller than a typical nearest-neighbor distance.
%
Moreover, the function itself and its derivatives up to order $n$ vanish continuously at the outer cutoff $r_\textrm{c}$.
It is demanded to be one order higher at the inner cutoff, because artifacts arise not only when atoms or entire neighbor shells cut through $r_\textrm{c}$ but also through $r_\textrm{i}$. 
This choice is made because short-ranged potentials, their forces and curvatures tend to be larger at  the inner radius than at the outer cutoff so that more care is required at $r_\textrm{i}$ than at $r_\textrm{c}$ . 

Shifting potentials have been discussed in particular in regard to the Wolf summation~\cite{Wolf1999JCP,Fennell2006JCP}.
Wolf \textit{et al.} showed that cutting and shifting the Coulomb interaction is equivalent to placing a charge-balancing countercharge at $r_\textrm{c}$, thereby reproducing an important element of the Evjen summation~\cite{Evjen1932PR}.
Applying the shifting procedure to the damped Coulomb interaction arising in the real-space part of the Ewald summation~\cite{Ewald1921AP} rather than to the original Coulomb interaction, allows the convergence with increasing $r_\textrm{c}$ to be quickly reached, even when neglecting the non-zero-wavenumber contributions to the Fourier portion of the Ewald sum. 
While systematic errors in the Wolf summation cannot be made arbitrarily small with the same low computational cost as with other Coulomb interaction summation techniques, most notably the particle mesh Ewald method~\cite{Essmann1995JCP}, it may yet be interesting for a variety of reasons: 
It can be used (i) for quick prototyping, (ii) in Monte Carlo simulations, which, unlike molecular dynamics, does not benefit from the simultaneous update or thermalization of all degrees of freedom, and (iii) in conjunction with multiple-time stepping schemes~\cite{Tuckerman1992JCP}.
This is why cutoff functions in the context of the Wolf summation are also investigated.
This includes a discussion of how to effectively use the (modified) Wolf summations when it fails to converge.

\section{Background}

\subsection{Conventional cutoff functions}

In principle, any shifted potential can also be obtained with a cutoff function defined implicitly through $f_\textrm{c}(r) \equiv U_{\textrm{s}n}(r)/U(r)$, where $n$ is the largest-order derivative of the potential going continuously to zero at the cutoff.  
The resulting cutoff function would not be near unity at a typical nearest-neighbor distance unless $r_\textrm{c}$ were very large. 
This is why shifting procedures should be generally inferior to more general cutoff functions with similar behavior for $U^{(n)}(r \lesssim r_\textrm{c})$. 
% whose derivative has similar scaling at $r_\textrm{c}$ as the S$n$ potential, but which reflects interactions at nearest-neighbor distances accurately. 
%
We can therefore dismiss simple shifting procedures as a competitive alternative to well designed shifting functions.

One of the most frequently used cutoff functions, supposedly proposed by Tersoff~\cite{Tersoff1986PRL}, is given by
\begin{equation}
f_\textrm{cF}(r) = \Theta(r_\textrm{i}-r) + 
\frac{
\Theta(r-r_\textrm{i}) \Theta(r_\textrm{c}-r)}{2}  
\left\{ 1 + \cos\left( \pi \frac{r-r_\textrm{i}}{r_\textrm{c}-r_\textrm{i}} \right) \right\} 
\end{equation}
This function, just like SF potentials, makes the force go linearly to zero as $r$ approaches $r_\textrm{c}$ but has a discontinuity in the curvature. 
Since $f_\textrm{cF}(r)$ is mirror symmetric about $(r_\textrm{m},1/2)$, where $r_\textrm{m} = (r_\textrm{i}+r_\textrm{c})/2$ can be called the mid-point, 
it has the same non-analyticity at $r_\textrm{i}$ and $r_\textrm{c}$. 

An improved version of and thus replacement for the SP potential can be generated with the cutoff function
\begin{equation}
f_\textrm{cP}(r) = \Theta(r_\textrm{i}-r) + 
\Theta(r-r_\textrm{i}) \Theta(r_\textrm{c}-r)
\sin\left( \frac{\pi}{2} \frac{r_\textrm{c}-r}{r_\textrm{c}-r_\textrm{i}} \right),
\end{equation}
which obeys the proposed rule of the cross-over function being one order more continuous at $r_\textrm{i}$ than at $r_\textrm{c}$.
This rule is meant to be useful when a potential decays with $1/r^3$ or faster, because the relative number of interactions, inner versus outer radius, where discontinuities in derivatives matters, scales with $(r_\textrm{i}/r_\textrm{c})^2$ in three spatial dimensions. 
For an electrostatic monomer-dipole interaction, the same order discontinuity at $r_\textrm{i}$ and $r_\textrm{c}$ would be recommended as this would balance errors at the inner radius and the outer cutoff. 
In lower dimensions, the exponents have to be corrected appropriately. 
% Additional cutoff functions for short-range potentials will be introduced and investigated in the results section. 

Other cutoff functions exist~\cite{Muser2022Ax}, e.g., functions that also have mirror symmetry about $(r_\textrm{m},1/2)$  but with higher-order vanishing derivatives at $r_\textrm{i}$ and $r_\textrm{c}$ than $f_\textrm{cF}$.   
However, they are not considered here, because they violate our mantra that more care needs to be taken at the inner than at the outer radius, and/or, because they have more ``structure'' than simple polynomials. 
Finally we note that this mantra does not apply to long-range potentials, where it may be beneficial to have smaller errors at $r_\textrm{c}$ than at $r_\textrm{i}$, in particular if $r_\textrm{i}=0$ is chosen. 

\subsection{Polynomial cutoff functions}

\begin{figure}[hbtp]
\centering
\resizebox*{10.0cm}{!}{\includegraphics{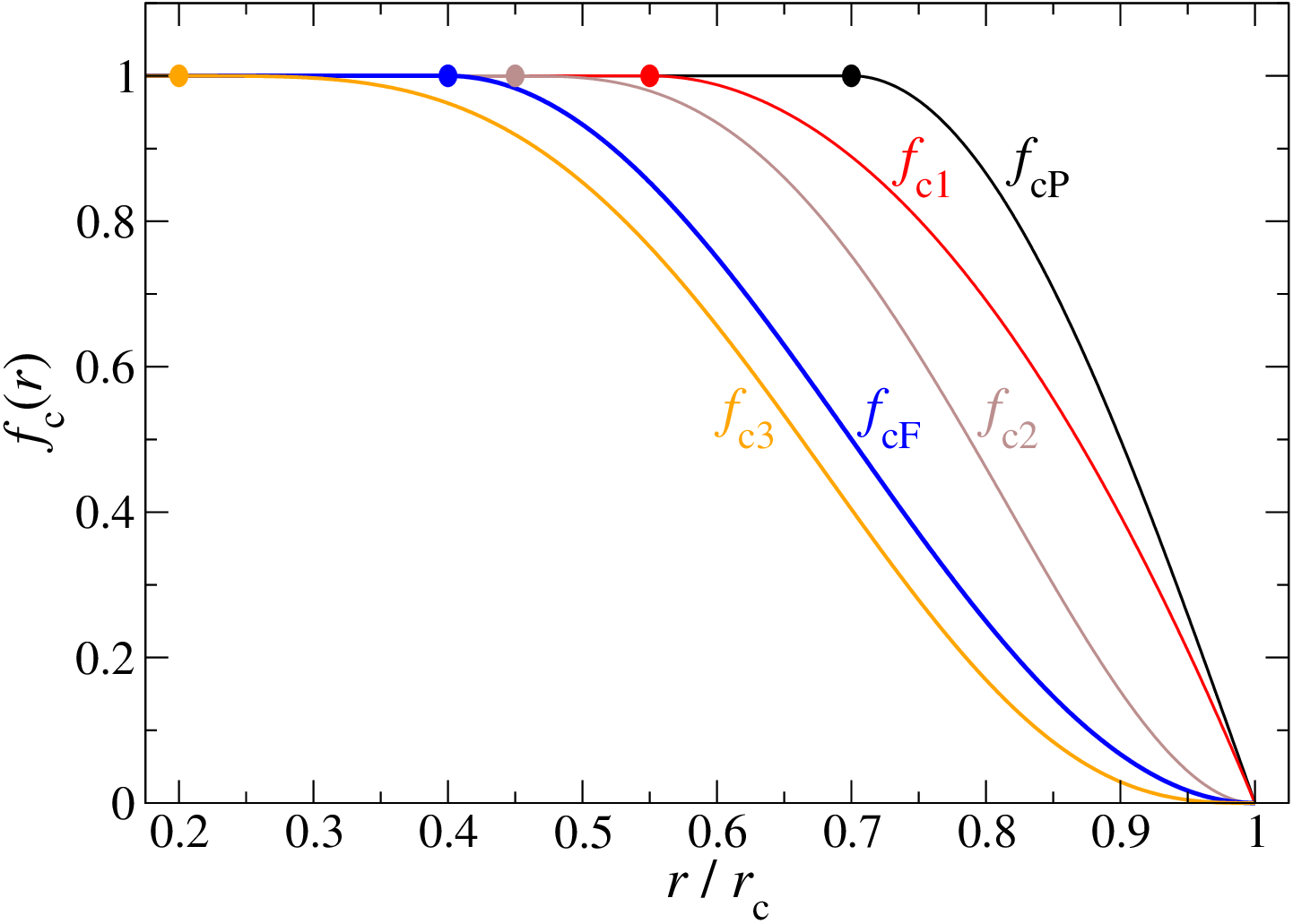}}
\caption{Selected cutoff functions.  Circles indicate the location of the inner cutoff radius $r_\textrm{i}$.
} \label{fig:cutOffFct}
\end{figure}

Cutoff-function proposed in prose in abstract and introduction are given by $f_{\textrm{c}n}(r) = 1 - P_n(x)$ with $x = (r-r_i)/(r_\textrm{c}-r_\textrm{i})$ and
\begin{subequations}
\begin{eqnarray}
P_1(x) & = & x^2 \nonumber\\
P_2(x) & = & 4 x^3 - 3 x^4 \nonumber\\
P_3(x) & = & 15 x^4 -24 x^5 + 10 x^6.
\end{eqnarray}
\end{subequations}
The $P_n(x)$ are constructed as the lowest-order polynomials to vanish with order $0,...,n+1$ at $x = 0$  and to assume $P(1) = 1$ while all derivatives up to order $n$ vanish at $x = 1$.
The resulting cutoff function are depicted in Fig.~\ref{fig:cutOffFct} together with $f_\textrm{cP}$  and $f_\textrm{c2}$.
Inner cutoffs were chosen for aesthetic reasons so that different functions do not cross. 

We are confronted with the task of determining ``optimal'' values for $r_\textrm{i}$, which depends on the cutoff function in addition to the (pair) potential and the property of interest.
Since the parametrization of a potential is done w.r.t. the cutoff function, it seems in place to suggest a generic choice in the \textit{one-size-fits-all} spirit.
Choosing $r_\textrm{i} = 0$ and $r_\textrm{i} = r_\textrm{c}$ are certainly lower and upper bounds.
However, they are obviously anything but helpful. 
One way of proceeding would be to demand that forces or derivatives at $r \ge r_\textrm{i}$ should not be greater in magnitude than at $r=r_\textrm{i}$.
For a dispersive $1/r^6$ attraction, $r_\textrm{i}/r_\textrm{c}$ would then turn out to lie within 0.8049 and 0.8221 for all cutoff functions discussed so far, except for the popular $f_\textrm{cF}(r)$ cutting function, which would require $r_\textrm{i}/r_\textrm{c}=0.7447$.
This smaller ratio arises because the non-analyticity of $f_\textrm{cF}(r)$ at the inner cutoff is as significant as at the outer cutoff.
Thus, larger compromises would have to be made on the cohesive energy using $f_\textrm{cF}(r)$ than for the remaining cutoff functions. 
Unfortunately, the just reported $r_\textrm{i}/r_\textrm{c}$ ratios still turn out too aggressive for the Lennard-Jones potential: 
the equation of state of an ideal,  face-centered cubic (fcc) Lennard-Jones remains discontinuous.
This undesired behavior could be eliminated by reducing the ratio to $r_\textrm{i}/r_\textrm{c}=2/3$.

\subsection{Cutoff functions infinitely often differentiable at $r_\textrm{c}$}

\begin{figure}[hbtp]
\centering
\resizebox*{8.5cm}{!}{\includegraphics{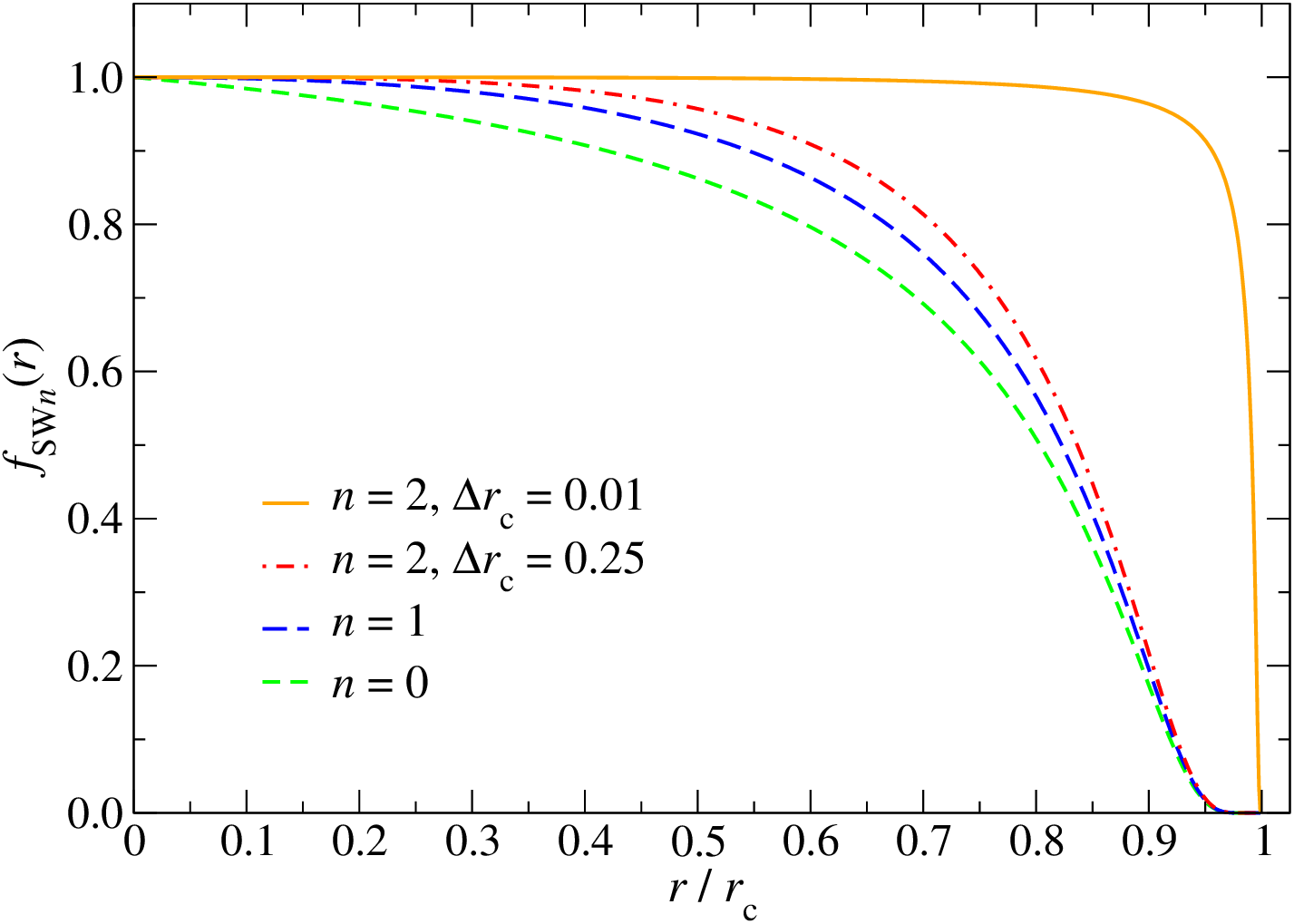}}
\caption{Selected Stillinger-Weber inspired cutoff functions. All dashed lines use $\Delta r_\textrm{c} = 0.25\,r_\textrm{c}$
} \label{fig:cutOffSW}
\end{figure}

Cutoff functions going to zero such that all their derivatives vanish at $r_\textrm{c}$ can be beneficial, 
e.g., in the context of generalized embedded-atom-method (EAM) based potentials, in which derivatives of the charge density enter the definition of the potential~\cite{baskes1989PRB,Jalkanen2015MSMSE}. 
This can be achieved with a function combining the Stillinger-Weber (SW)~\cite{Stillinger1985PRB} cutoff function and the idea of a polynomial expansion pursued in this paper.
Specifically, 
\begin{equation}
\label{eq:eamCutOffFunction}
f_{\textrm{SW}n}(r) = \frac{\mathcal{T}[\textrm{denom}(r),0,n]}{1+\exp\left( \frac{\Delta r_\textrm{c}}{r_\textrm{c}-r} \right)} \Theta(r_\textrm{c}-r),
\end{equation}
is such a cutoff function.
Here, $\textrm{denom}(r)$ is the denominator of the quotient on the r.h.s. of the equation,  ${\mathcal{T}}[...]$ denotes a Taylor series expansion as above, and $\Delta r_\textrm{c}$ determines, as a function of $n$, how closely to  $r_\textrm{c}$ the cutoff function assumes the value 0.5. 
Some selected SW generalized cutoff functions are shown in Fig.~\ref{fig:cutOffSW}.
The original one proposed by Stillinger and Weber corresponds to $n = 0$.

For $\Delta r_\textrm{c}\ll r_\textrm{c}$ and/or large values of $n$, the effect of $f_{\textrm{SW}n}(r)$ will be similar to that of a harsh cutoff, as $f_\textrm{c}(r)$ is close to unity up to the immediate vicinity of $r_\textrm{c}$, in which case the disadvantages from harsh cutoffs are inherited.
We see no reason to use $f_{\textrm{SW}n}(r)$ for regular short-range potentials, however, it could benefit, for example, the systematically modified embedded atom method,~\cite{Jalkanen2015MSMSE} for which the embedding energy depends on (higher-order) derivatives of the embedding density.
Higher-order derivatives place higher demands on the way in which the charge density is brought down to zero at $r_\textrm{c}$, in particular when an individual atom breaks its final bond. 

Although using $f_{\textrm{SW}n}$ is disadvised for regular short-range potentials, it appears to be a suitable candidate to cut off long-range potentials, since its analyticity is of higher order at $r_\textrm{c}$ than at $r_\textrm{i}$.
This expectation is explored in the context of the Wolf summation in Sect.~\ref{sec:Wolf}. 

\section{Cutting short-range potentials}

The generic (pair) potential used to describe non-bonded interaction is the Lennard-Jones (LJ) potential
\begin{equation}
U(r) = U_0 \left\{ \left( \frac{r_0}{r} \right)^{12}
- 2 \left( \frac{r_0}{r} \right)^6 \right\},
\end{equation}
where $U_0$ is the binding energy of a dimer and $r_0$ its equilibrium bond length. 
The standard cutoff used for LJ is $r_\textrm{c} = 2.5\,\sigma$, where
$\sigma = \sqrt[6]{2}\,r_0$ is also called the LJ radius. 
Often, the LJ potential is merely shifted using this default value.
This procedure is standard practice and certainly acceptable.
Nonetheless, simultaneous improvements on both accuracy and computing time should be possible, which is explored next. 

To demonstrate the effect of the various cutting schemes, the fcc LJ crystal will be investigated. 
It allows artifacts to be highlighted, while keeping computing times and numerical errors minimal.
The local structure certainly differs between LJ crystals with well-defined neighbor shells and liquid Lennard-Jonesium, which is close to random-sphere packing. 
As a consequence, typical bond distances at zero or what-would-be ambient pressure are less than $r_0$ in the crystal but greater $r_0$ in the liquid.
Next-nearest neighbor distances and associated coordination numbers, to be defined, e.g., through  a skew-normal-distribution analysis of peaks in the radial distribution $g(r)$~\cite{Sukhomlinov2017JCP},  differ even more between crystal and liquid. 
Including into the discussion the radial distribution functions arising in (united-atom based) models of polymers makes it even more difficult to identify guidelines for how to pick $r_\textrm{i}$ and $r_\textrm{c}$ so that they both coincide with minima in $g(r)$. 
Thus, any final choice should yield robust results no matter how $r_\textrm{i}$ and $r_\textrm{c}$ relate to the maxima and minima in $g(r)$ of any particular system of interest. 
Any critical situation is included in the analysis when analyzing the cohesive energy and the equation of state (EOS) in the range $0.85 \le a_0/r_0 \le 1.3$, where $a_0$ is the (mean) nearest-neighbor bond length. 
This is because the energy of an individual LJ bond is already positive at $r = 0.85\,r_0$, which is a situation of a very high compressive stress or force.
At the other end at $a_0 = 1.3\,r_0$, a LJ bond can be considered broken, because this bond length is beyond the inflection point of the LJ potential, i.e., past the point of maximum tensile force.

The goal is to identify parameters for the cutoff function(s) and radii that globally outperform the standard cut-and-shift procedure.
To this end, we chose arbitrarily $r_\textrm{c} = 2.5\,r_0$, which reduces the interaction volume to 70\% compared to that of the default cutoff, $r_\textrm{c} = 2.5\,\sigma$,  and thereby the number of force evaluations by a similar percentage.
Of course, it would be a simple matter to include mean-field corrections for the cohesive stress~\cite{intVeld2007JCP} so that smaller cutoff radii could be trivially achieved without losing accuracy. 
However, they would not be useful for heterogeneous systems, e.g., when surfaces are present.
Moreover, such corrections are not always available in popular software packages. % as they occur, for example, when a soft part of a medium is compressed by a stiff constituent.
This is why mean-field corrections are not included in this study either. 

\begin{figure}[hbtp]
\centering
\subfloat[Cohesive energies $U_0$ of various approximation schemes for fcc Lennard-Jonesium.]
{\resizebox*{6.7cm}{!}{\includegraphics{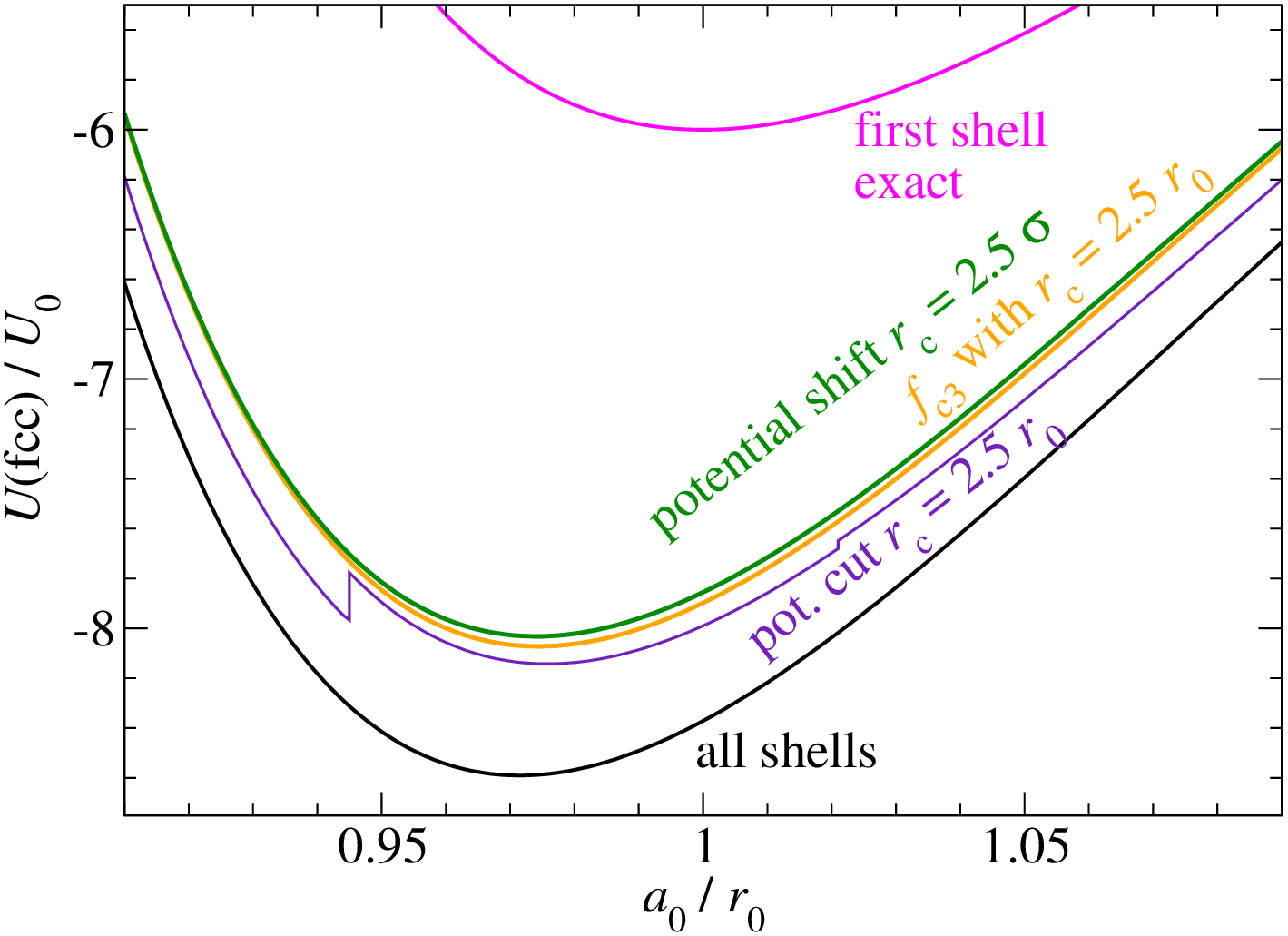}}}\hspace{5pt}
\subfloat[Weighted errors of various approximation schemes to the exact cohesive energy.]
{\resizebox*{7.1cm}{!}{\includegraphics{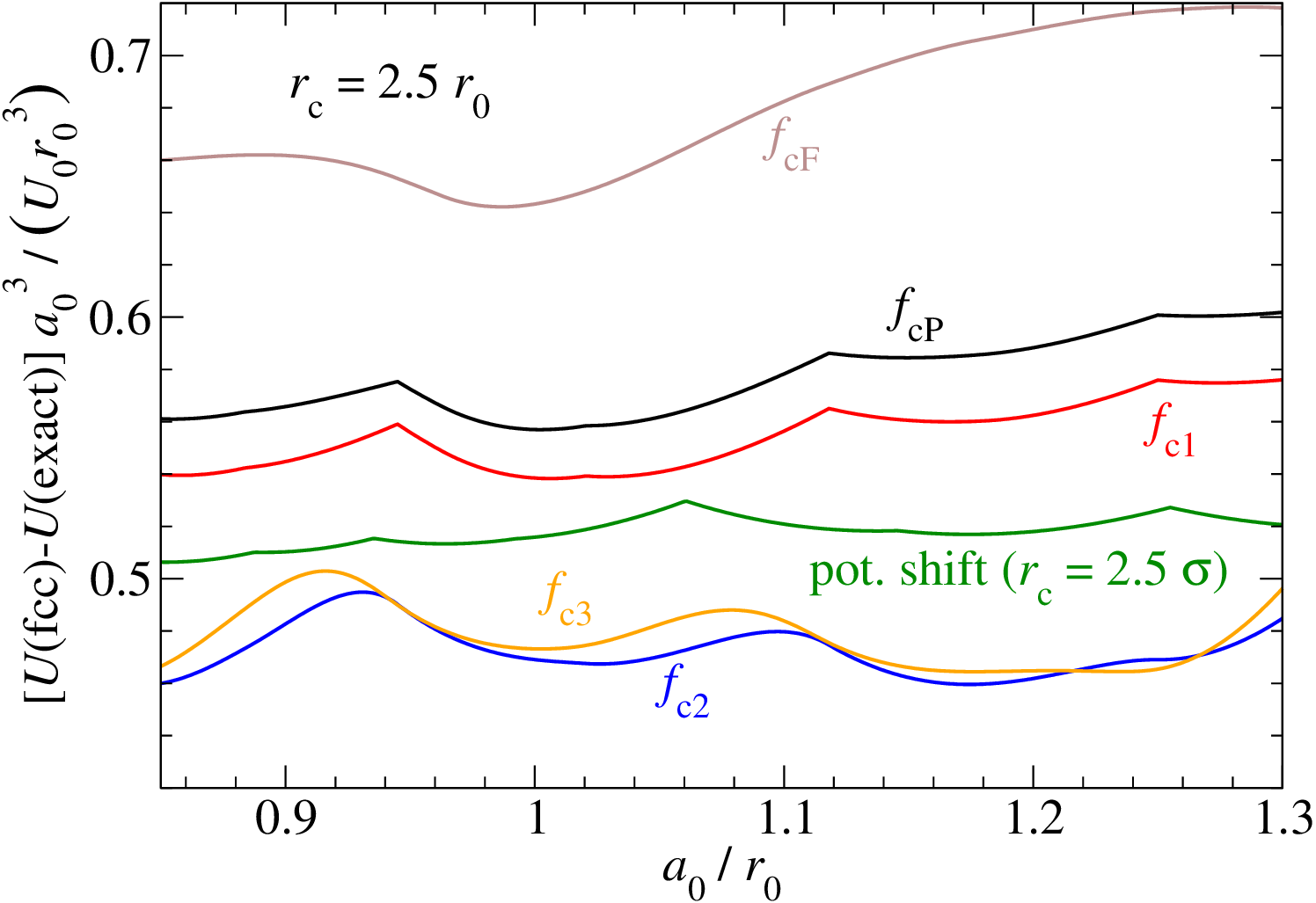}}}\hspace{5pt}
\vspace{2pt}
\subfloat[Global equation of state.]
{\resizebox*{7.0cm}{!}{\includegraphics{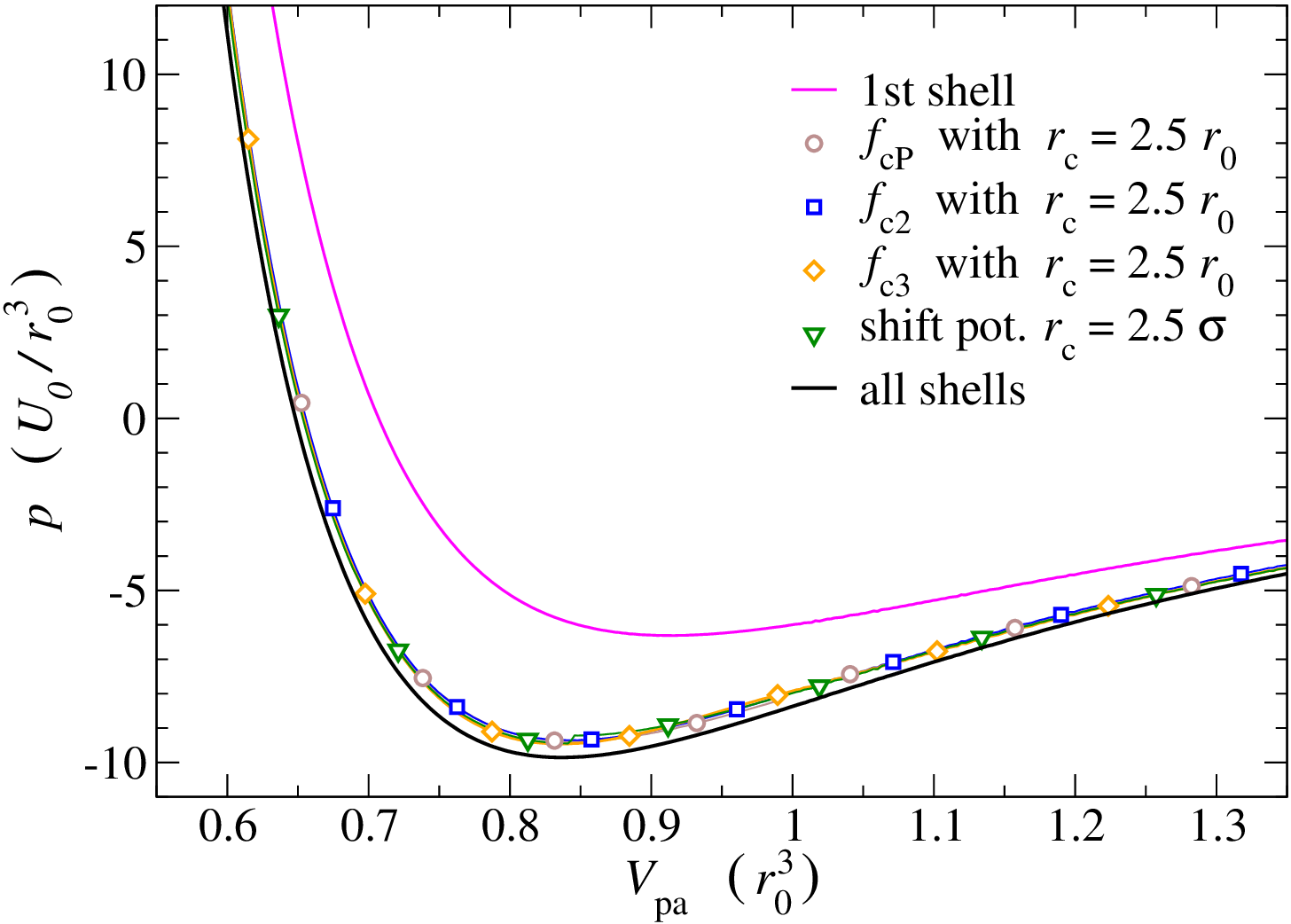}}} \hspace{5pt}
\subfloat[Zoom into equation of state.]
{\resizebox*{7.0cm}{!}{\includegraphics{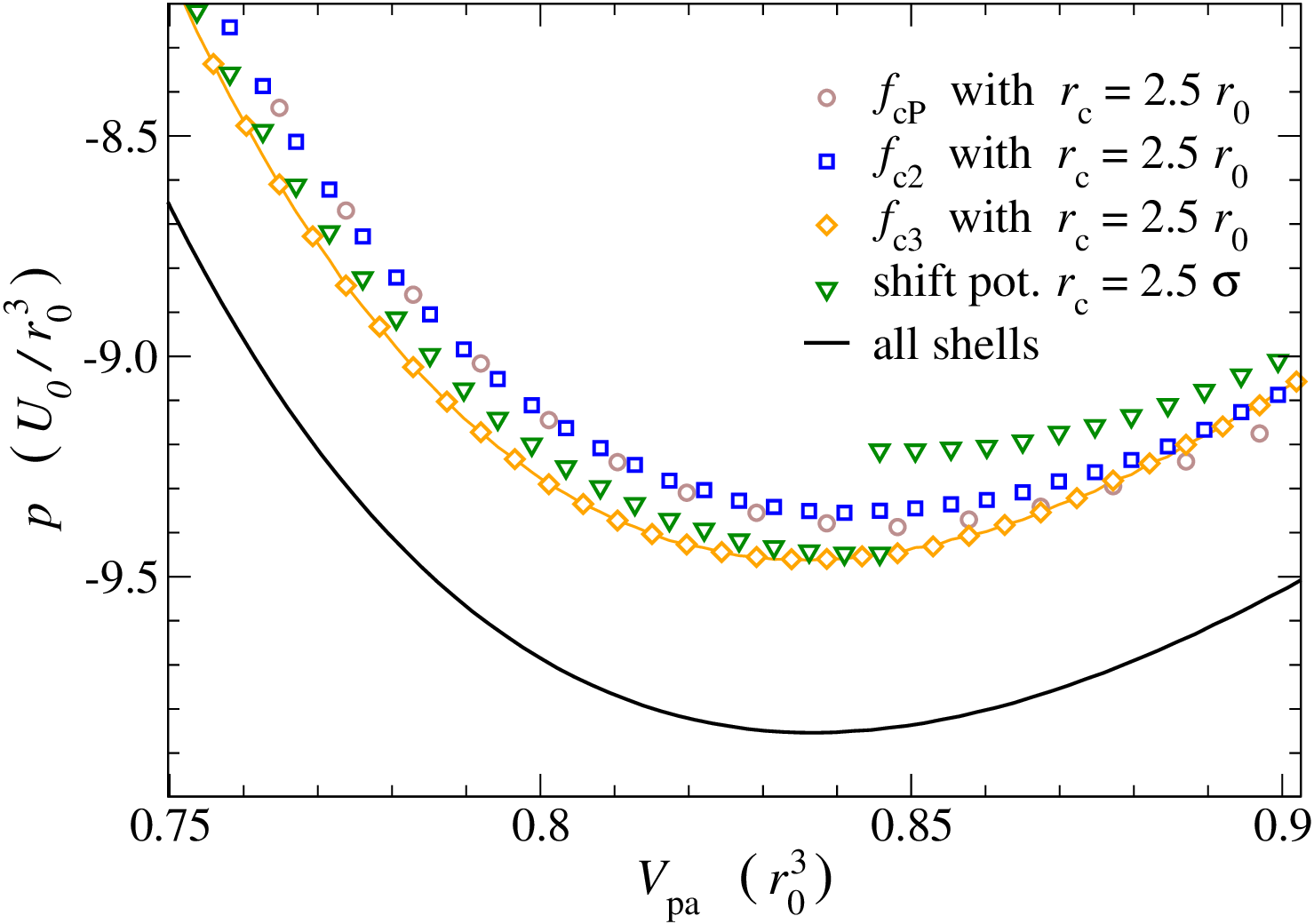}}} \hspace{5pt}
\caption{Effect of cutting procedure on energies as function of nearest-neighbor distance $a_0$ and equation of state.  } 
\label{fig:energFCC}
\end{figure}

For $r_\textrm{c} = 2.5\,r_0$, an inner cutoff of $r_\textrm{i} = 2\,r_\textrm{c}/3$ was found benefical.
It makes the cohesive energy of an fcc crystal be just below the default cut-and-shift procedure with $r_\textrm{c} = 2.5\,\sigma$, at least in the ``interesting range'' of $0.85 < a_0/r_0 < 1.3$, which is demonstrated in Fig.~\ref{fig:energFCC}, where the values of the pertinent potential energies are almost within line width in panel (a). 
Both the default cut-and-shift as well as the $f_{\textrm{c}3}(r)$ cutting yield a similar minimum in the cohesive energy of about $8~U_0$, per atom, while the nearest-shell approximation yields exactly $6~U_0$.
% , since the coordination number is $Z = 12$ and each atom ``owns'' each bond half. 
%
The exact binding energy is about $8.59~U_0$. 
Fig.~\ref{fig:energFCC}(a) also reveals that using a harsh, unshifted cutoff at $2.5~r_0$ does not significantly lower the energy compared to a method using the same cutoff radius but the high-order smoothing function $f_\textrm{c3}(r)$. 
However, the discontinuities occurring when using harsh cutoffs generally yield unacceptable behavior. 

Fig.~\ref{fig:energFCC}(b) resolves the error $\Delta U = U(\textrm{appr.})-U(\textrm{exact})$ over a relevant range.
Errors are multiplied with $a_0^3$ to make results approximately constant. 
Values turn out close to the ones expected from the mean-field correction to the dispersive interaction, i.e., 
\begin{eqnarray}
\frac{ \Delta U_\textrm{disp}}{U_0} & \approx & \int_{r>r_\textrm{c}} \!\!\! \mathrm{d}^3r\,\rho \,\left( \frac{r_0}{r} \right)^6 
 =  \frac{ \sqrt{32}\pi}{3} \frac{r_0^6}{r_\textrm{c}^3\,a_0^3} ,\nonumber
\end{eqnarray}
the numerical prefactor evaluating to approximately 0.38 after having inserted the fcc atomic number density of $\rho = \sqrt{2}/a_0^3$.
Since repulsion reduces the binding energies, 0.38 is merely a lower bound for the numbers reported in Fig.~\ref{fig:energFCC}b. 

Since the energies of the various approximations schemes are quite close to each other, so will be their EOS.
In fact, they turn out to be within line width in Fig.~\ref{fig:energFCC}c, except for the nearest-neighbor approximation revealing a significantly reduced (theoretical) maximum cohesive stress.
However, zooming into parts of the EOS resolves that the standard cut-and-shift procedure induces a discontinuous EOS. 
Similar discontinuities also occur under compression, however, their relative effect is negligible.
Of course, even minor thermal fluctuations smear out the discontinuities so that one certainly does not need to be concerned when using the standard $r_\textrm{c} = 2.5\,\sigma$ LJ cut-and-shift procedure.
Nonetheless, they can become relevant for other potentials or for smaller cutoffs.

It can be summarized that the $f_{\textrm{c}2}(r)$ and $f_{\textrm{c}3}(r)$ cutoff function lead to smaller errors than the standard cut-and-shift procedure for the cohesive energy and the EOS in the range what we deemed to be interesting.
At the same time, they require only about 70\% of the force evaluations.
However, this latter point is only advantages when look-up tables for interatomic potentials and forces are used.
Otherwise, the additionally required floating point operations needed to evaluate forces from smoothly cut potentials would be prohibitively expensive.

\section{Wolf summation}
\label{sec:Wolf}

\subsection{Background}

Ewald~\cite{Ewald1921AP} demonstrated that Coulomb interactions in periodically repeated systems can be meaningfully summed up by dividing the summation into a real-space and a Fourier or reciprocal-space contribution. 
The latter containts two terms at zero wave vector, one of which is the so-called self-interaction energy and the other the electrostatic energy resulting from the electrostatic field generated by the mean dielectric polarization. 
In detail, given a charge-density distribution of $\rho(\mathbf{r}) = \sum_i q_i \delta(\mathbf{r}-\mathbf{r}_i)$ with zero net charge, the electrostatic energy reads~\cite{Ewald1921AP,deLeeuw1980PRS}
\begin{eqnarray}
U_\textrm{C}(\{\mathbf{r}\}) & = & 
\sum_{i,j>i}  \frac{q_i q_j}{4\pi\varepsilon_0}
\frac{\mathrm{erfc}(k_\textrm{E}r_{ij})}{r_{ij}} - 
\sum_i \frac{q_i^2}{4\pi\varepsilon_0} \frac{k_\textrm{E}}{\sqrt{\pi}}
\nonumber\\
& & + \frac{\mathbf{p}^2_\textrm{tot}}{2(2\varepsilon_\textrm{ext}+1)\varepsilon_0 V} 
% \sum_i \left\vert q_i (\mathbf{r}_-\mathbf{r}_i^0) \right\vert^2
+ \frac{1}{2\varepsilon_0 V} \sum_{\mathbf{k},k\ne 0} \frac{\left\vert \tilde{\rho}(\mathbf{k})\right\vert^2}{k^2} 
\, e^{{-{k^2}}/{4k_\textrm{E}^2}},
\label{eq:Ewald}
\end{eqnarray}
where $V$ is the volume of the periodically repeated (simulation) cell, $\tilde{\rho}(\mathbf{k})$ is the Fourier transform of $\rho(\mathbf{r})$, and $\mathbf{p}_\textrm{tot}$ the total dipole moment of the simulation cell, i.e., $\mathbf{p}_\textrm{tot} = \sum_i q_i (\mathbf{r}_i-\mathbf{r}_{0i})$ assuming the dipole moment for the set of reference coordinates $\{\mathbf{r}_0\}$ to vanish.
Finally, $\varepsilon_\textrm{ext}$ is the relative permittivity of an external embedding medium.
% , e.g.,  $\varepsilon_\textrm{ext}=1$ for vacuum and $\varepsilon_\textrm{ext}=\infty$ for metals. 
%
Including its effect matters for finite clusters, which are placed into a simulation cell with a vaccuum buffer, in which case $\varepsilon_\textrm{ext}=1$.

The last summand on the r.h.s. of Eq.~\eqref{eq:Ewald} can become irrelevant for sufficiently small $k_\textrm{E}$.
This might have enticed Wolf \textit{et al.}~\cite{Wolf1999JCP} to ignore that term completely. 
In order to effectively enforce charge-neutrality within $r_\textrm{c}$, Wolf \textit{et al.}~\cite{Wolf1999JCP} used a cut-and-shift potential, and corrected the self-interaction energy to 
\begin{equation}
U_\textrm{DSP}^\textrm{self}(i) = -
\left( \frac{k_\textrm{E}}{\sqrt{\pi}}
+\frac{\mathrm{erfc}(k_\textrm{E}r_\textrm{c})}{2r_\textrm{c}} \right)
\frac{q_i^2}{4\pi\varepsilon_0} .
\end{equation}
Although simply cutting and shifting potentials is problematic for reasons discussed above as well as in Refs.~\cite{Patra2003BJ,Fennell2006JCP,Waibel2019JCTC}, this original Wolf summation is taken as the reference for alternative cutting procedures investigated here below.

Picking $k_\textrm{E}$ properly when performing a real Ewald summation is crucial to achieve a good compromise between speed and accuracy.
Using a fast Ewald method, $k_\textrm{E}$ can be kept constant irrespective of the system size, or, particle number $N$.
For the conventional Ewald sum, the apparently optimum choice is $k_\textrm{E} \propto 1/\sqrt[6]{N}$, in which case the computational effort to yield results with a target error scales as $N^{3/2}$, both in real as well as in reciprocal space~\cite{Perram1988MP,Frenkel2002}.
For both fast and conventional Ewald summation, increasing the demand on accuracy by a given factor then only necessitates an increase in computing time scaling sub-logarithmically in this factor. 

The large convergence rate of Ewald summations cannot be achieved using the Wolf summation.
However, even an algebraic dependence would be desirable, in particular in the context of Monte Carlo simulations, which, unlike molecular-dynamics simulation,  does not profit from a parallel update of all degrees of freedom. 
To ensure convergence using the Wolf summation, $k_\textrm{E}$ must be made a function of $r_\textrm{c}$.
As discussed in more detail here below, the overall best choice when using a Wolf summation turns out to satisfy $k_\textrm{E} \approx 1/\sqrt{a_0 r_\textrm{c}}$, where $a_0$ is a typical distance of adjacent anions and cations. 

\subsection{Convergence for ideal and perturbed rock-salt structures}
\label{sec:rocksalt}

\begin{figure}[hbtp]
\centering
\subfloat[Absolute values.]{%
\resizebox*{6.9cm}{!}{\includegraphics{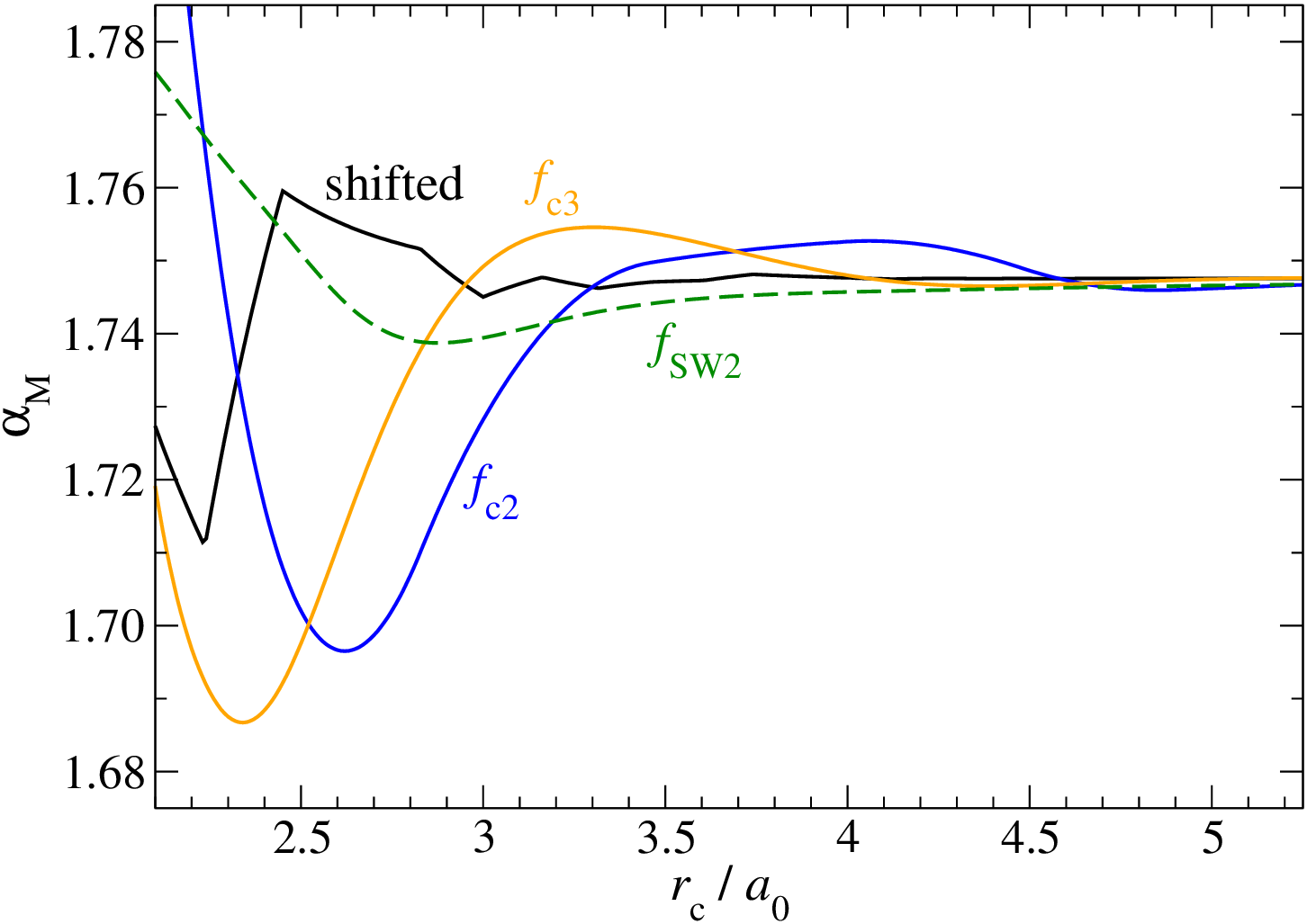}}}\hspace{5pt}
\subfloat[Magnitude of deviation.]{%
\resizebox*{6.8cm}{!}{\includegraphics{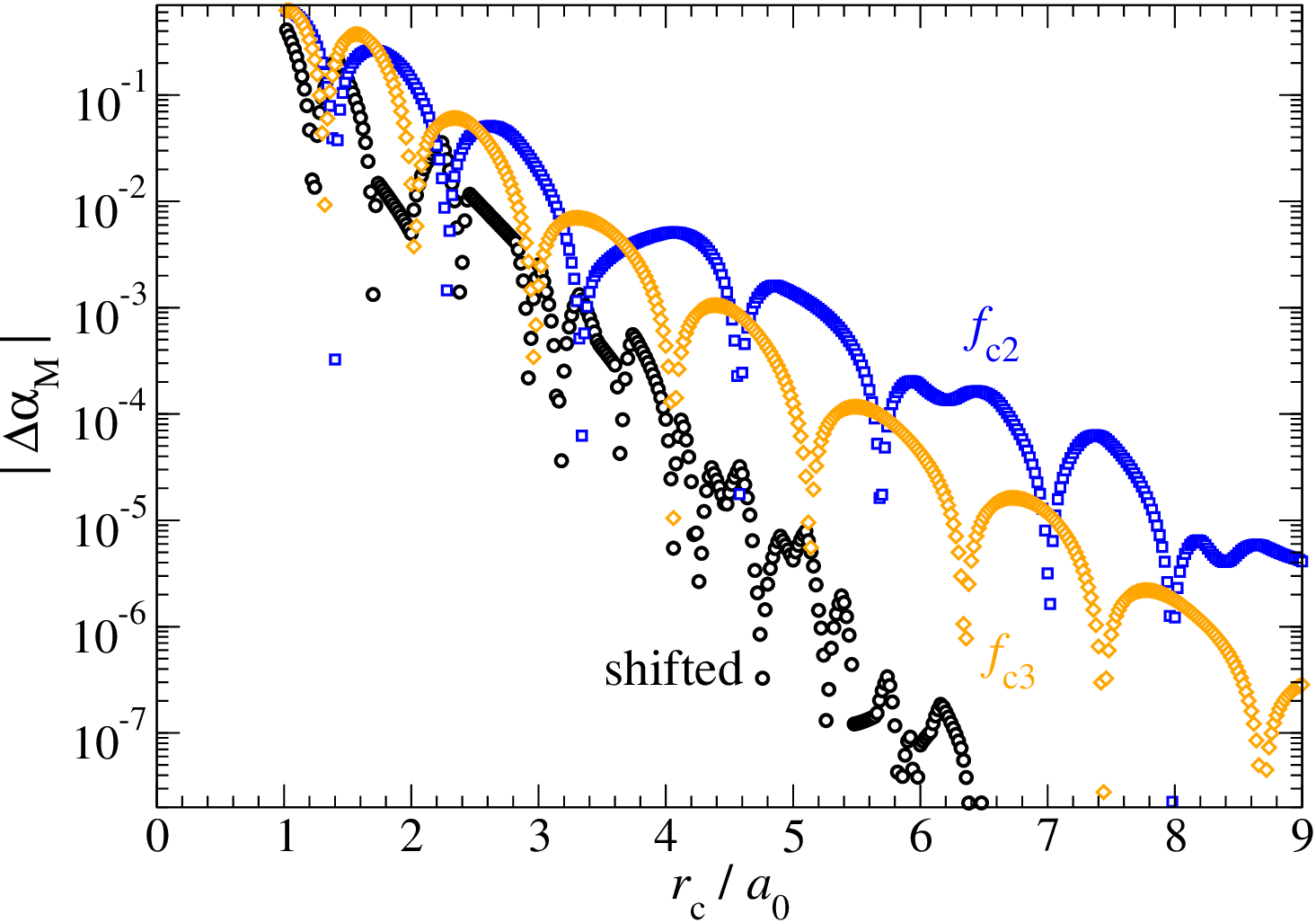}}}
\caption{Convergence of Madelung constant $\alpha_\textrm{M}$ in ideal rock salt with increasing cutoff radius $r_\textrm{c}$ in units of the bond length $a_0$ for various cutting schemes for the damped Coulomb interaction. 
} \label{fig:cutOffFctC}
\end{figure}

A first convergence analysis for the Wolf summation is presented in Fig.~\ref{fig:cutOffFctC}. 
Panel (a) shows the Madelung constant $\alpha_\textrm{M}$, while panel (b) depicts the magnitude of its error. 
The latter reveals that convergence is not only algebraic but even exponential with $r_\textrm{c}$ for this highly symmetric structure when using  $k_\textrm{E} = 1/\sqrt[4]{a_0^3 r_\textrm{c}}$.
% , where $a_0$ is the nearest-neighbor spacing between anions and cations in the rocksalt structure.
%
While the original Wolf summation converges the most quickly, kinks in $\alpha_\textrm{M}$ indicate indirectly that the original summation will unavoidably lead to artifacts.
Essentially exponential convergence is also obtained for $f_{\textrm{c}2}$ and $f_{\textrm{c}3}$, for which $r_\textrm{i}=r_\textrm{c}/2$ was used.  
Although rates are a little less than for the original Wolf summation, the new summations are much better behaved whenever a neighbor sell cuts through the cutoff radius. % , which is why we would consider them to be the better choice for simulations. %
The $f_{\textrm{SW}2}$-data was obtained using $r_\textrm{i}=0$ and $\Delta r = r_\textrm{c}$. 

Before proceeding to less idealized cases, some observations will be reported. 
First, ideal rock salt was the only structure for which choosing $k_\textrm{E} \approx 1/\sqrt[4]{a_0^3 r_\textrm{c}}$ was clearly optimal.
In all other cases, $k_\textrm{E} \approx 1/\sqrt{a_0 r_\textrm{c}}$ turned out to be the apparently best option for reasons stated further below.
Second, rock salt was the only structure for which $f_{\textrm{c}2}$  and $f_{\textrm{c}3}$ ``outperformed'' $f_{\textrm{SW}2}$ at large $r_\textrm{c}$.
Third, making the order $n$ at $r_\textrm{i}$ greater than at $r_\textrm{c}$, i.e.,  replacing $1-P_n(x)$ in the definition of $f_{\textrm{c}n}(r)$ with $P_n(1-x)$, did not improve results.
Fourth, $f_\textrm{SW2}$ outperformed all other $f_{\textrm{SW}n}$.
Observations 3 and 4, whose reasons we do not yet understand, also hold for the other investigated structures.

Deviations from the ideal rock-salt structure were also investigated.
First, a small random distance was added to each atomic coordinate so that the far field of an atoms is identical to that of a point charge augmented with a random dipole. 
Second, the charge of each atom on an ideal lattice was augmented or reduced randomly by half an elementary charge with the constraint that the net charge remains unchanged. 
The result is an ionic solid solution with positional disorder. 
The such produced configurations will be called random-dipole and random-charge crystals, respectively.

For sufficiently large systems, the Madelung constants of both random crystals is identical to that of regular rock salt.
This is because (a) the field of a random dipole or higher-order multipole has a random direction so that placing another multipole into its field has, on average, zero potential energy and (b) the expectation value of the product $q_i q_j$ satisfies $\langle q_i q_j \rangle = \langle q_i \rangle \langle q_j \rangle$ in the limit of infinite particle numbers.
In finite systems, systematic deviations occur because the fluctuation of a given charge is perfectly correlated with that of its periodic images but slightly anti-correlated with all other charges and their periodic images.

\begin{figure}[hbtp]
\centering
\subfloat[Plus symbols and crosses show the original and a $f_{\textrm{SW}2}$-modified Wolf  summation, respectively. The Ewald paramaeter was chosen as 
$k_\textrm{E} = 1.2/(a_0^{1-\beta} r_\textrm{c}^\beta)$ with $\beta = 1/3$ (blue) and $\beta = 1/2$ (red). \label{fig:randDipole}
% 
% $k_\textrm{E} = 1.2/{a_0r_\textrm{c}}^\beta$ 
% with $\beta = 1/3$ (blue) and $\beta = 1/2$ (red)]. % The data were fit with $f(r_\textrm{c}) = f(\infty)+b r_\textrm{c}^\gamma$ and the asymptotic values drawn as dashed lines.
]{%
\resizebox*{6.9cm}{!}{\includegraphics{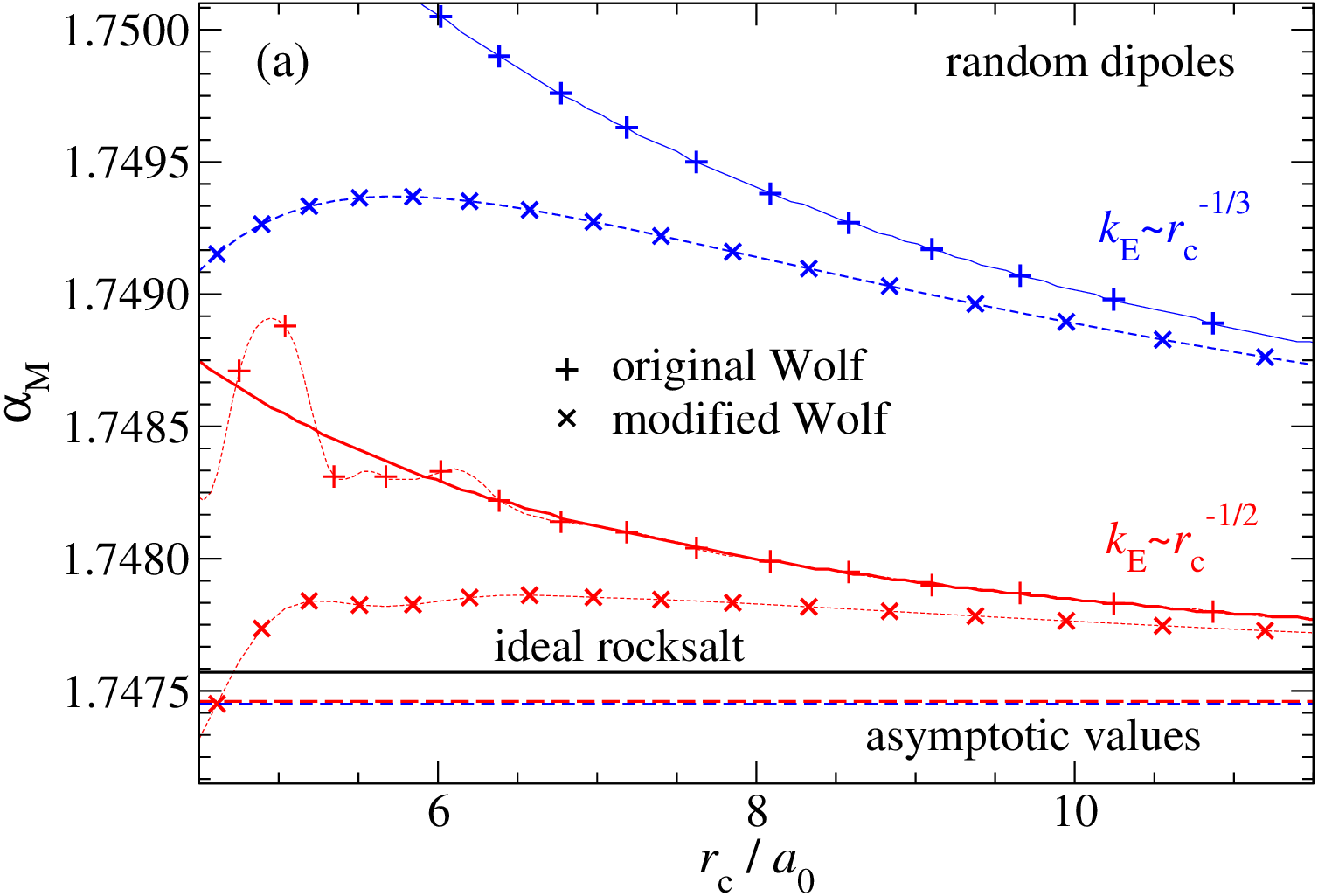}}}\hspace{5pt}
\subfloat[\label{fig:randCharge} Different symbols refer to different prefactors $\kappa$ used for $k_\textrm{E} = \kappa/\sqrt{a_0r_\textrm{c}}$, i.e., $\kappa=1.5$ (brown triangles), $\kappa=1$ (red circles) and $\kappa = 0.5$ (blue squares). Orange crosses show the Evjen summation.]{%
\resizebox*{6.8cm}{!}{\includegraphics{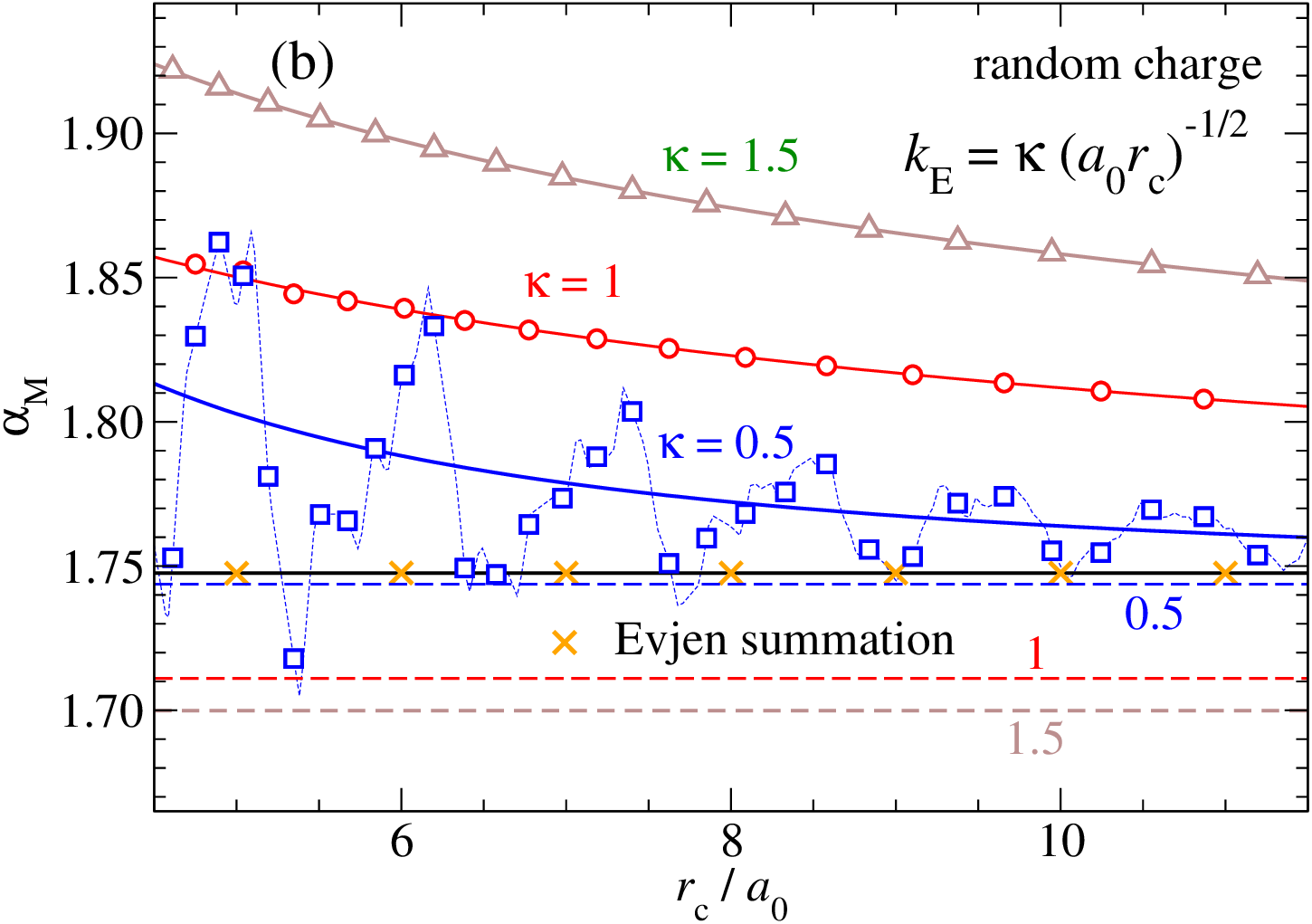}}}
\caption{Convergence of the mean Madelung constant $\alpha_\textrm{M}$ for (a) random-dipole and (b) random-charge crystals containing $24\times24\times 24$ atoms in total. For $r_\textrm{c}>8\, a_0$, the Wolf sums were fit with 
$\alpha_\textrm{M} = \alpha_\textrm{M}(\infty) + \beta/ r_\textrm{c}^\gamma$ and shown as colored, solid lines on the entire domain. 
The asymptotic values, $\alpha_\textrm{M}(\infty)$, are drawn as dashed lines in their respective color.
In both panels, the Madelung constant of ideal rock salt is drawn as a solid, black line.
} \label{fig:randomCrystal}
\end{figure}

Fig.~\ref{fig:randDipole} shows that the Wolf summation converges to the proper effective or mean Madelung constant $\alpha_\textrm{M}$ for the random-dipole crystal (within statistical fluctuations from one random realization to the next) if $k_\textrm{E}$ is made an appropriate function of $r_\textrm{c}$, e.g., 
$k_\textrm{E} = \kappa/(a_0 r_\textrm{c})^{\beta}$ with the prefactor $\kappa = 1.2$ and exponents $\beta = 1/2$ or $\beta = 1/3$.
This is not surprising, since the dipole-dipole interaction of oriented dipoles is just no longer integrable in three spatial dimensions, so that sums over randomly oriented dipoles are unconditionally integrable or summable. 
Fig.~\ref{fig:randDipole} allows the following, additional observations to be made:
The Wolf-summation results can be fit quite accurately using 
$\alpha_\textrm{M} = \alpha_\textrm{M}(\infty)+\beta/r^\gamma $
at large $r_\textrm{c}$.
A smaller exponent $\beta$ leads to a smaller exponent $\gamma$, however, the asymptotic scaling is reached at smaller $r_\textrm{c}/a_0$ ratios. 
In the given example, $\gamma = 1$ for $\beta = 1/3$ and $\gamma = 3/2$ for $\beta = 1/2$.
Moreover, the modified Wolf summation has the same asymptotic approach to $\alpha_\textrm{M}(\infty)$ as the original summation.

It is also noted that the SW2-modified Wolf summation tends to be closer to the exact result than the original Wolf summation, however, asymptotic scaling sets in at larger $r_\textrm{c}$. 
Besides producing continuous forces and potential curvatures at $r_\textrm{c}$, this is one reason why the use of the  SW2-modified Wolf summation would be suggested for simulations for which $r_\textrm{c}$ is a fixed quantity. 
However, extrapolating $\alpha_\textrm{M}(r_\textrm{c}\to\infty)$ is more easily done using the original Wolf summation, or, for example, the $f_{\textrm{c}3}$-modified Wolf summation. 
For this reason, most of the subsequent convergence analysis is made on the original Wolf summation, 

For the random-charge crystal, the Wolf summation no longer converges to the correct Madelung constant, as is revealed in Fig.~\ref{fig:randCharge}, at least as long as $r_\textrm{c}$ is less than half the size of the periodically repeated cell. 
This time, the prefactor $\kappa$ to the $k_\textrm{E} = \kappa/ \sqrt{a_0 \, r_\textrm{c}}$ was varied.
The exponent $\gamma$ in the (seemingly) asymptotic $\alpha_\textrm{M} = \alpha_\textrm{M}(\infty)+\beta/r^\gamma$ relation was again not universal but turned $\gamma \lesssim 1/2$ for $\kappa \gtrsim 1$. %
Thus, being locally charge neutral in a stochastic sense, is not a sufficiently strong condition for the Wolf summation to converge.
If positive and negative charges separate deterministically, which can be caused by a structural heterogeneity on scales exceeding $r_\textrm{c}$, the Wolf summation will obviously be even more erroneous than for random charge neutrality. 
%
% As will be discussed in the conclusions, it may yet be useful in the context of Monte Carlo simulations. 
%
% This is one reason why its convergence will be explored further for situations in which the order of the local multipoles ranges between those considered so far. 

% Combining the ideas of the ``Fermi cutoff function'' with the one by Stillinger and Webber:
% \begin{equation}
% (1.+\exp(a))/(1+\exp(a/(1-x)))
% \end{equation}

\subsection{Convergence for crystalline and liquid silica}

The convergence rate of the modified Wolf summation is also explored on crystals of lower symmetry than rock salt and a corresponding ionic melt, namely silica.
As reference crystal, cristobalite was chosen.
It is a polymorph of silica, in which the silicon atoms form a cubic diamond lattice and the bridging oxygen atoms predominantly rotate in a safe distance about their average, crystallographic  positions, which are located half way between two adjacent silicon atoms~\cite{Dove1995MM,Muser2001PCM}.
The (local) symmetry of atoms in this polymorph is lower than in rock salt, because the field gradients on oxygen atoms even in the crystallographic positions are unisotropic, while the anisotropy of fields of atomic positions in the ideal rock salt structure appears first in its third spatial derivative. 
Since the real positions of oxygen are quite distant from the crystallographic ones, oxygen atoms tend to sit at sites with a relatively rather large electric field. 
Silica is simulated with the potential proposed by van Beest, Kramer, and van Santen (BKS)~\cite{vanBeest1990PRL} using a house-written code described before~\cite{Muser2001PCM}.  
Despite some shortcomings, the BKS potential has reproduced various properties of liquid~\cite{Vollmayr1996PRB} and crystalline~\cite{Muser2001PCM,Herzbach2005JCP} silica.

\begin{figure}[hbtp]
\centering
\resizebox*{8.5cm}{!}{\includegraphics{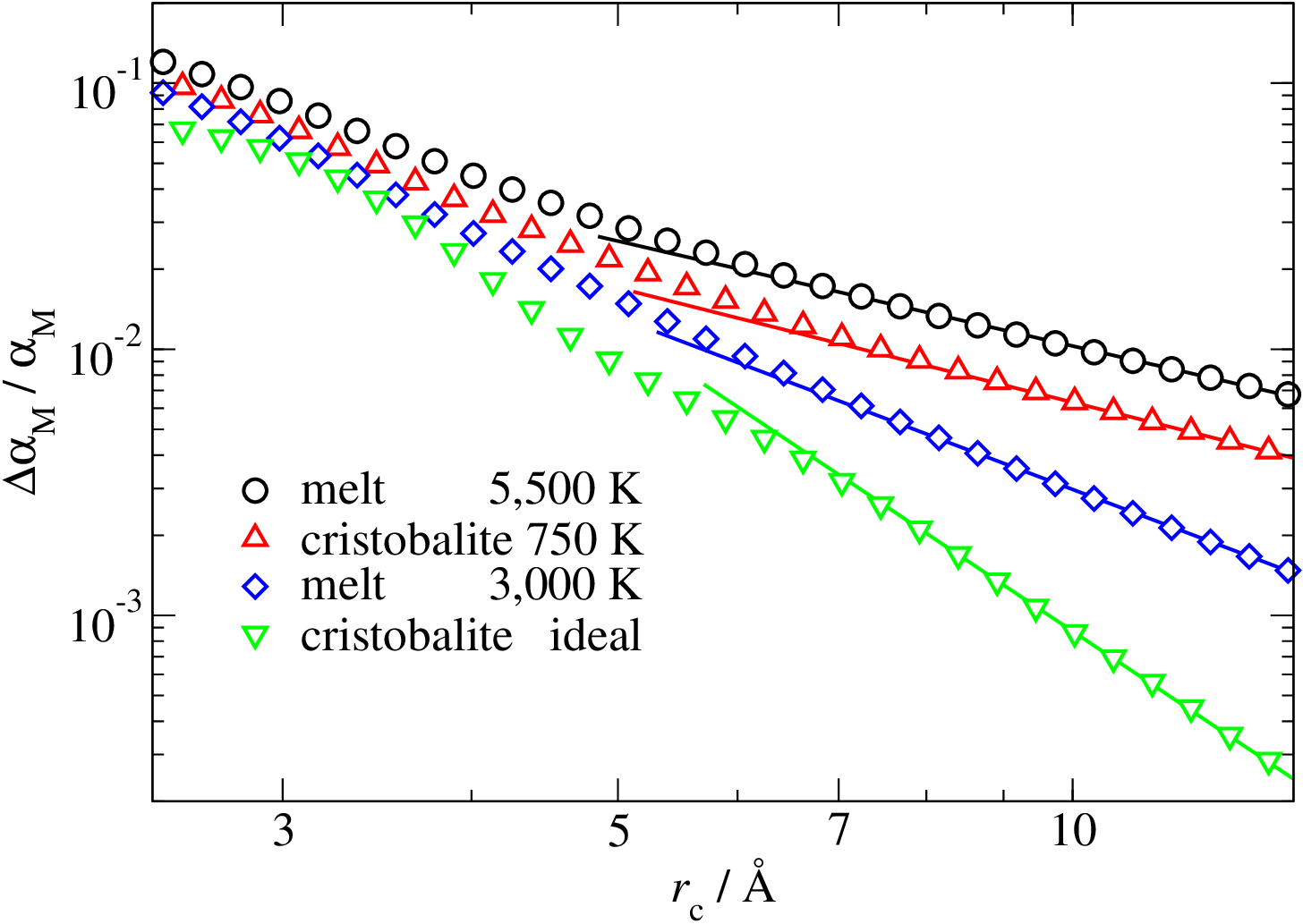}}
\caption{\label{fig:compSiO2}
Relative error in the Coulomb energy $\Delta \alpha_\textrm{M}/\alpha_\textrm{M}$ as a function of the cutoff radius $r_\mathrm{c}$ in units of {\AA} for a modified Wolf summation. The studied systems were a melt at 5,500~K (black circles) and at 3,000~K (blue diamonds) as well as a thermal cristobalite crystal at 750~K (red triangles up) and an ideal crystal at the same density, for which, however, all atoms were placed onto their ideal lattice positions (green triangles down). Lines represent powerlaws with exponents $\gamma = 1.31$, 1.43, 2.15, 3.8 (top to bottom). 
} 
\end{figure}

Fig.~\ref{fig:compSiO2} shows the relative error in the Coulomb energy, which was obtained for silica melts at two different temperatures as well as for cristobalite, one time with oxygen atoms being constrained to their crystallographic positions and one time at a temperature just above the phase transformation temperature from the high-symmetry $\beta$-cristobalite phase to the optically active $\alpha$-cristobalite~\cite{Schmahl1992ZK}. 
As expected, the Wolf summation converges more quickly for the ideal, crystallographic crystal than for the thermal crystal, for which the Wolf summation converges similarly quickly, or, depending on viewpoint, slowly as the random-dipole crystal considered in Sect.~\ref{sec:rocksalt}. 

\begin{figure}[hbtp]
\centering{%
\resizebox*{6.9cm}{!}{\includegraphics{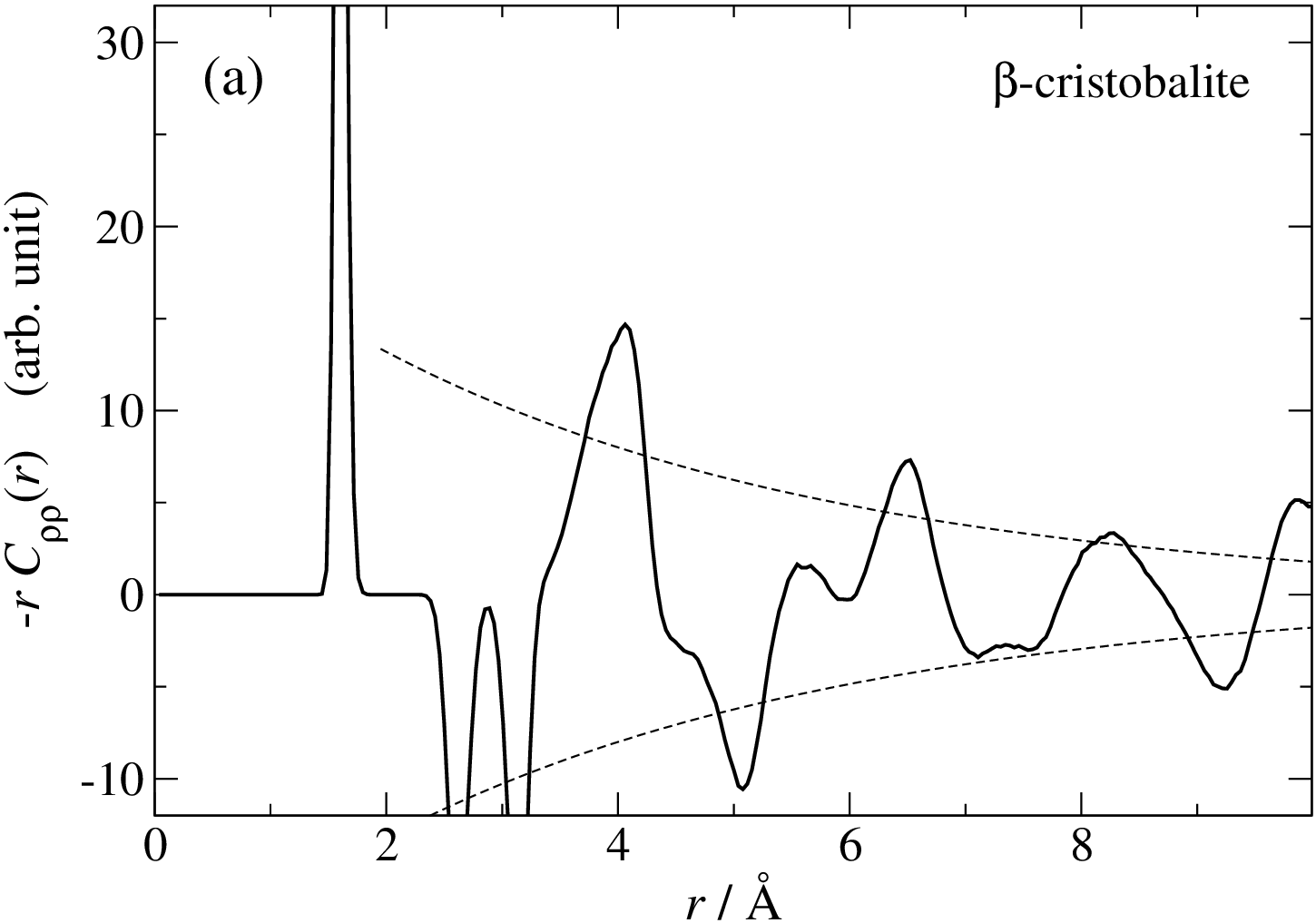}}}\hspace{5pt}
{%
\resizebox*{6.8cm}{!}{\includegraphics{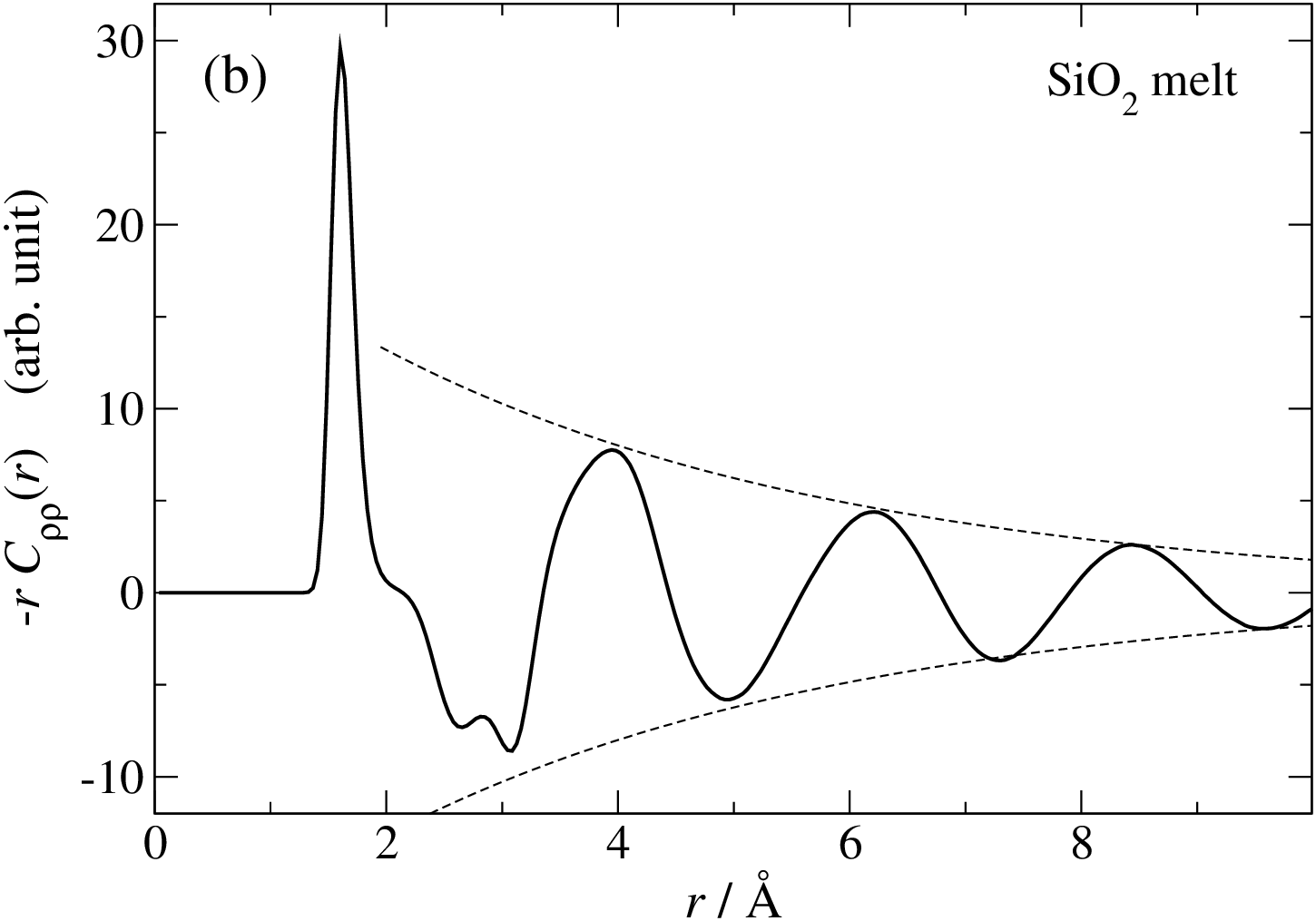}}}
\caption{Weighted charge-density autocorrelation function, $-r C(r)$, as a function of distance $r$ for (a) $\beta$-cristobalite at 750~K and (b) a SiO$_2$ at 3,000~K. The dashed lines are exponential function proportional to $\pm \exp(-r/\zeta)$ with $\zeta = 4$~\AA. 
} \label{fig:OZ}
\end{figure}

A surprising result of Fig.~\ref{fig:compSiO2} is the relatively fast convergence of the Wolf summation for the ``low-temperature'' ($T =3,000$~K) melt, which is not only faster than at $T =5,500$~K melt but also faster than for the 750~K, thermal crystal. 
This may have to do with the fact that Madelung sums should actually converge for homoegeneous melts since the (partial) density autocorrelation function in dense liquids are damped oscillations at large $r$ not ``suffering'' from distant neighbor shells carrying large number of atoms and thereby preventing lattice sums from unconditional convergence.
Ultimately, (twice) the electrostatic energy per point charge can be cast as an integral over the charge-density autocorrelation function, $C_{\rho\rho}(r) \equiv \langle \rho(0) \rho(r)\rangle$ via
\begin{eqnarray}
U_\textrm{C} & = & \frac{1}{4\pi\varepsilon_0} \int_{0^+}\!\mathrm{d}^3 r\, \frac{C_{\rho\rho}(r)}{r} = \frac{1}{\varepsilon_0}\int_{0^+}^\infty \!\mathrm{d}r \, r \, C_{\rho\rho}(r)
\label{eq:energyOZ}
\end{eqnarray}
where $0^+$ is meant to indicate that self-interactions of charges at $r = 0$ are excluded from the integral. 
The (negative) integrand in the last term of Eq.~\eqref{eq:energyOZ} is shown in Fig.~\ref{fig:OZ}.
It reveals that subsequent peaks in the integrand become ever smaller in the melt but not nexessarily in the crystal. 
In dense, three-dimensional liquid this behavior can be rationalized using the Ornstein-Zernike theory~\cite{Baxter1970JCP}, which predicts density oscillations to obey asymptotically $\cos(r/\lambda+\varphi)\exp(-r/\zeta)$, where $\lambda$ is a wavelength, $\varphi$ a phase shift and $\zeta$ a correlation length. 
Maxima and minima in the negative integrand, which could be interpreted as a Madelung constant density, are located near the maxima of the partial dislike and like-ion radial distribution functions, respectively.  
While the magnitude of the integrand for large $r$ is clearly bound by a simple exponential in the liquid, this is not true for the crystal, where the extrema at $r \gtrsim 9$~{\AA} and $r \lesssim 10$~{\AA} are more pronounced than those in the interval $7$--$8$~\AA. 

As a small side remark to this article, we wish to note that the computational burden of the regular Ewald summation can be slightly reduced when the Fourier part of the interaction is not evaluated every time step but only every ${\cal{O}}(a_0 k_\textrm{E})$ time steps.
Such a reduction is possible, because the long-wavelength dynamics are slower than the ones associated with short wavelengths.
The CPU time needed for the real-space sum would then scale as before with $N r_\textrm{c}^3$, while the Fourier part would be reduced from $k_\textrm{E}^3 N^2$  to $k_\textrm{E}^4 N^2$, assuming that, say a wavenumber cutoff of $k_\textrm{c} \approx 4 k_\textrm{E}$ is generally acceptable. 
For systems with stark (charge) heterogeneity on arbitrary wavelengths, it would be required to chose $k_\textrm{c} \propto 1/r_\textrm{c}$ as to avoid uncontrollable summation errors, which would otherwise arise if a structural heterogeneity existed on a wavelength exceeding simultaneously $r_\textrm{c}$ and $2\pi/k_\textrm{c}$.
Minimizing the total CPU time through a proper choice of $k_\textrm{E}$ would then lead to a $N^{10/7}$ rather than a $N^{3/2}$ scaling of the numerical effort with particle number $N$. 

\section{Discussion and conclusions}
In this article, the search of the proper balance between accuracy and efficiency when cutting potentials was discussed.
This is certainly an important, albeit somewhat neglected issue.
The need for its discussion was recognized while writing a review on interatomic potentials~\cite{Muser2022Ax}, where it would have been inappropriate to suggest new cutoff functions and their properties. 

This article emphasizes that well-designed cutoff functions should generally outperform cut-and-shift potentials and that the discontinuities in cutoff functions at the inner radius deserve at least the same attention as at the outer cutoff radius, in particular for short-range potentials decaying more quickly than $1/r^3$. 
On the simple Lennard-Jones potential, it shows that the standard cutting procedure can be optimized in that computing time (when using tabulated potentials and forces) and errors on energy could be reduced by roughly 20 to 30\%. 
% accurate results on energies can be obtained while reducing computing times as well as cutoff artifacts in the equation of state and related properties. 
%
While these gains are relatively minor, the incredibly large number of computations assuming Lennard-Jones potentials might make it worth while implementing the cutting-off procedure defined in this work. 

Although the Wolf summation~\cite{Wolf1999JCP} was scrutinized in earlier work~\cite{Fennell2006JCP,Cisneros2013CR}, we could not deduce from it a clear message of the conditions when it converges and when it fails and how to best pick the Ewald parameter $k_\textrm{E}$. 
Here, we found that it is well behaved for most homogeneous systems but that local charge neutrality must be obeyed more systematically than in a purely stochastic sense. 
Moreover, we found $k_\textrm{E} = 1/\sqrt{a_0 r_\textrm{c}}$ as a kind of optimum choice in that it  worked well for all investigated practical situations involving thermal or structural fluctuations. 
In fact, this choice appears is the ``sweet spot'', similar to a critically damped case, where for the more general choice of  $k_\textrm{E} = \kappa/(a_0^{1-\beta} r_\textrm{c}^\beta)$, the scaling of $\alpha_\textrm{M}$ with $r_\textrm{c}$ crosses over from an ``overdamped'' (convergence with small exponent) to oscillatory behavior upon either an increasing $\beta$ or decreasing $\kappa$.

Despite being problematic when charge density is not strictly locally neutral, the Wolf summation can still be useful under such conditions.
However, an exact summation of the $k$-space contribution would have to be made sporadically, in particular in Monte Carlo simulations, which, unlike molecular dynamics, does not benefit from a simultaneous update of all propagated degrees of freedom.
The entire simulation between two such $k$-space evaluations would then constitute one large trial move so that the latest configuration after many steps using only the Wolf summation would be considered a trial configuration. 
It could be accepted or rejected using, for example, in the Metropolis algorithm~\cite{Metropolis1953JCP}, where the energy difference between new and old $k$-space contribution, $\Delta U_k$, would enter the Boltzmann factor. 
While the rejection of such a time-intensive trial move is certainly regrettable, a reasonable scaling of the overall numerical effort with particle number $N$ should be achievable. 
Of course, as is the case with the traditional Ewald summation as used in molecular dynamics~\cite{Perram1988MP}, $r_\textrm{c}$ would have to increase algebraically with $N$ so that the absolute error induced by local ``Wolf moves'' decreases with increasing $N$. 

\section*{Acknowledgement(s)}
MHM acknowledges helpful discussion with Sergey Sukhomlinov, Lars Pastewka, and Joshua Weißenfels. 

\bibliographystyle{unsrt}
% \bibliographystyle{apacite}
% \bibliography{cutoff}

\end{document}